\documentclass{emulateapj}

\slugcomment{Accepted for publication in Astrophysical Journal}
\shorttitle{M51 Star Cluster Burst} \shortauthors{Hwang \& Lee}

\begin{document}

\title{Vestige of the Star Cluster Burst in M51\altaffilmark{*}}

\author{\sc Narae \ Hwang$^{1,2,}$\altaffilmark{3} and Myung Gyoon \ Lee$^2$}
\affil{$^1$ National Astronomical Observatory of Japan \\2-21-1 Osawa
Mitaka, Tokyo 181-8588, Japan} 
\affil{$^2$ Astronomy Program, Department of Physics and Astronomy,\\
Seoul National University, Seoul 151-747, Korea}

\altaffiltext{*}{Based on observations made with the NASA/ESA {\it Hubble
Space Telescope}, obtained from the data archive at the Space Telescope
Science Institute, which is operated by the Association of Universities for
Research in Astronomy, Inc., under NASA contract NAS 5-26555.}
\altaffiltext{3}{Japan Society of the Promotion of Science Research Fellow}

\begin{abstract}
We present a study of the star cluster formation in M51 based on the image
data taken with Advanced Camera for Surveys (ACS) and Wide Field Planetary
Camera 2 (WFPC2) of Hubble Space Telescope (HST). We have derived the
star cluster formation rate using the ages and masses of about 2,000 star
clusters estimated by comparing photometric data ($F336W$, $F435W$,
$F555W$, and $F814W$) with theoretical population synthesis models. The
star cluster formation rate increased significantly during the period of $100
\sim 250$ Myr ago. This period roughly coincides with the epoch of dynamical
encounters of two galaxies, NGC 5194 and NGC 5195, expected by theoretical
models. The age distribution of the star clusters also shows two peaks at
about $100$ Myr and $250$ Myr ago. The star cluster mass ranges from
$10^3$ to $10^6 M_{\odot}$ and the mass function can be represented by
a power law with an index ranging from $\alpha = -2.23 \pm 0.34$ for
$t<10$ Myr to $\alpha = -1.37 \pm 0.11$ for $t>100$ Myr. The mass
function of star clusters older than $10$ Myr also appears to display the
steepest distribution with $\alpha \approx -1.50$ at around $200$ Myr ago,
near the expected epoch of the galaxy interaction. We also confirm the
correlations of cluster size increasing with cluster mass (with a best fit slope
of $0.16 \pm 0.02$), and with cluster age ($0.14 \pm 0.03$).
\end{abstract}


\section{Introduction}
\label{intro}

Star clusters are an excellent tool to probe the formation and evolution of
galaxies due to their high brightness enough to be detected even in some
nearby galaxies. They are considered to be composed of stars that formed at
the same time. Some star clusters are also known for their longevity: globular
clusters are as old as the Universe itself. Therefore, we can investigate the
very ancient star cluster formation events as well as the recent ones in a
galaxy by estimating the age of star clusters and constructing their age
distribution.

M51 is an interacting galaxy system that is composed of a Sbc galaxy (NGC
5194) and a fainter SB0 galaxy (NGC 5195). M51 is an optimal and interesting
target for the study of star cluster formation for several reasons. (1) It is
abundant with interstellar media (see a recent study by Schuster et al. 2007
and references therein) that serve as basic ingredients for stars and star
clusters. HII region studies (e.g., Scoville et al. 2001, Lee, J. H. 2009 in
preparation) showed that current formation of stars and star clusters is
actively going on in M51. (2) There are some theoretical \citep{too72,sal00a}
and observational \citep{dur03} studies suggesting that the two galaxies in
M51 experienced a single or multiple encounters some hundred Myr ago.
Therefore it is possible to find any correlations between these dynamical
events and the star formation events in M51. (3) M51 is located at a distance
of 8.4 Mpc \citep{fel97} so that, with the resolving power of the Hubble
Space Telescope ($HST$), it is possible to resolve, at least, some star clusters
in M51.

Several studies used the $HST$ data to survey star clusters in M51.
\citet{bik03} detected about 1000 star clusters in a small region near the
center of M51 and found that the cluster formation rate decreases with age
during the period of $10 - 1000$ Myr ago. However, they found no evidence
for an increased cluster formation rate around $200 - 400$ Myr ago, when
the first encounter of NGC 5194 and NGC 5195 was expected. \citet{bas05a}
showed that the cluster formation rate increased about $50 - 70$ Myr ago
when the second (or last) passage of NGC 5195 was expected by a multiple
passage model \citep{sal00a}. Still, no clear sign of increased star cluster
formation around the $200 - 400$ Myr ago was shown by \citet{bas05a}. A
hint on the increased cluster formation rate around this epoch was given by
\citet{lee05}. They derived the age distribution of about 400 resolved clusters
in M51 using the $HST$ archival data and showed that, when compared with
another typical late type galaxy M101, there are an increased number of star
clusters with ages of $100 - 500$ Myr, which are consistent with the expected
epoch of the first encounter. However, their study did not cover the entire
area of M51 and the number of star clusters used in their study was small.

One limiting factor that has haunted previous studies on star clusters in M51 is
the limited field of view and the insufficient depth of the available data. It
became possible to reduce this problem, when the deep and wide field image
data taken with the Advanced Camera for Surveys (ACS) onboard the $HST$
were released \citep{mut05}. Now it enables to investigate the star clusters
concentrated not only in the disk of NGC 5194 but also those around the
outer halo including the companion galaxy NGC 5195 in a homogeneous
quality. This has led to several new studies on the star clusters in M51:
\citet{hwa06} on faint fuzzy clusters in NGC 5195, \citet{sch07} on the
cluster size distribution, \citet{hwa08} on the photometric properties of star
clusters, \citet{has08} on the cluster luminosity function, and \citet{sch09}
on the distribution of star cluster formation in the disk. However, there is not
yet any detailed study using this new data on the correlation between the star
cluster formation and dynamical interactions in M51, and it is not yet clear
how the star cluster formation was affected by the dynamical events in M51.

In this study, we investigate the age distribution of M51 star clusters and
derive the cluster formation rate using the catalog of the star clusters in M51
given by \citet{hwa08}. This star cluster catalog includes about 3,600 clusters
with $V_{F555W}<23$ mag and provides photometric data in $F435W$,
$F555W$, and $F814W$ bands. We compare the photometric data in this
catalog with the population synthesis models by \citet{bc03} to estimate the
age and mass of star clusters. A preliminary study of the age distribution of
M51 star clusters was already presented in \citet{hwa07}, implying a possible
correlation between the dynamical encounters of galaxies and the age
distribution of star clusters in M51. However, this result was based on
$F435W$, $F555W$, and $F814W$ band photometry data, and could be
severely affected by the age-reddening degeneracy in age estimation.
Therefore, we included the $F336W$ band image data of M51 taken with
$HST$ WFPC2 to improve the accuracy of age estimation for this study.

The organization of this paper is as follows: Section \ref{data} describes the
data. Section \ref{phot} briefly introduces the star cluster selection and
photometry procedures along with the reduction of $F336W$ band data.
Section \ref{sed} describes the age estimation method. We present the main
results including the age and mass distribution, and the star cluster formation
rate in Section \ref{result}. We discuss the primary results in Section
\ref{discuss}. Finally, a summary and conclusion is given in Section
\ref{sum}.

\section{Data}
\label{data}

\begin{figure}
\plotone{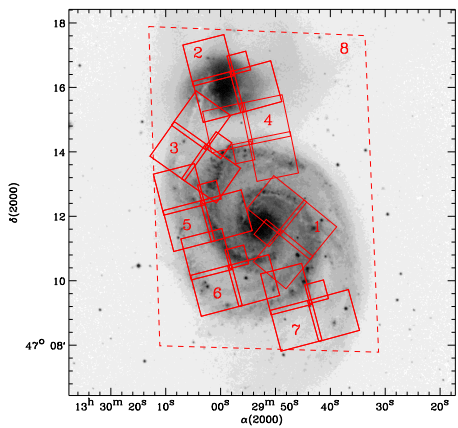} \caption{Locations of the fields with the $HST$
 ACS data (field \#8 in dashed line) and WFPC2 data (solid line) used in this study
 overlaid on a $12\arcmin \times 12\arcmin$ Digitized Sky Survey image of M51.
 Detailed information on each pointing is listed in Table~\ref{obssum}.
\label{ufov}}
\end{figure}

The main data used in this study are $F435W$, $F555W$, and $F814W$ band
images taken with $HST$ ACS through the Hubble Heritage program 10452
(PI: S. V. W. Beckwith). We also used the $F336W$ band images taken with
$HST$ $WFPC2$ for six fields through the observing program 10501 (PI: R.
Chandar), and another $F336W$ band image set that covered the central
region of NGC 5194, available in the $HST$ archive. A brief information on the
$F336W$ band data set is provided in Table \ref{obssum}. Figure \ref{ufov}
shows the observed fields overlaid on the $12\arcmin \times 12\arcmin$
Digitized Sky Survey (DSS) image of M51. Seven WFPC2 $F336W$ fields
cover a major part of M51, but their coverage is still much smaller than the
ACS field.

\begin{deluxetable*}{ccccccc}[b]
\label{obssum}
\setlength{\tabcolsep}{0.05in}
\tablecaption{A list of $HST$ WFPC2 $F336W$ band data for M51 used in this study\label{obssum}}

\tablehead{ \colhead{Field No.} & \colhead{Prop. ID} & \colhead{RA
(J2000)} & \colhead{Dec (J2000)} & \colhead{Instruments} &
\colhead{Filters} &
\colhead{Exposure Times} \\
 & & \colhead{[hh mm ss]} & \colhead{[dd mm ss]} & & & \colhead{[sec]} }
\startdata
    1 &  7375 &  13 29 47.81 & +47 11 32.84 & WFPC2 & $F336W$ &  2 x 600  \\
    2 & 10501 &  13 30 00.52 & +47 15 40.40 &       & $F336W$ &  2 x 1300 \\
    3 &       &  13 30 07.80 & +47 14 00.50 &       & $F336W$ &  2 x 1300 \\
    4 &       &  13 29 51.68 & +47 15 02.85 &       & $F336W$ &  2 x 1300 \\
    5 &       &  13 30 05.90 & +47 11 40.00 &       & $F336W$ &  2 x 1300 \\
    6 &       &  13 30 00.87 & +47 09 39.37 &       & $F336W$ &  2 x 1300 \\
    7 &       &  13 29 46.32 & +47 08 34.39 &       & $F336W$ &  2 x 1300 \\
\enddata
\end{deluxetable*}


Data reductions for ACS data were carried out by the STScI including
multi-drizzling and image combination. More detailed information on these
$HST$ ACS data reductions was given in \citet{mut05}.\footnote{See also
http://archive.stsci.edu/prepds/m51/.} All WFPC2 data that were retrieved
from the $HST$ archive were reduced using the ``on-the-fly'' calibration,
which automatically uses the best reference files for calibration. The WFPC2
data reduction steps include bad pixel masking, analog-to-digital correction,
bias and dark subtraction, and flat-field correction. There are two individual
images per each WFPC2 pointing. These two WFPC2 image data were
combined using the $combine$ task in STSDAS package of
IRAF\footnote{IRAF is distributed by the National Optical Astronomy
Observatory, which is operated by the Association of Universities for Research
in Astronomy (AURA) under cooperative agreement with the National Science
Foundation.} with CRREJECT option in order to eliminate cosmic rays and
increase the signal-to-noise ratio.

We adopted a distance of $8.4\pm0.6$ Mpc [$(m-M)_0=29.62$] to M51
determined from the planetary nebula luminosity function in M51
\citep{fel97}. The corresponding linear scale is 40.7 pc
arcsec$^{-1}$. The foreground reddening toward M51 is low,
$E(B-V)=0.035$, and the corresponding extinctions are $A_B=0.150$,
$A_V=0.115$, $A_I=0.067$, and $A_U=0.190$ mag \citep{sch98}.
The total magnitudes and colors are $B^T=8.96\pm0.06$ mag and
$(B^T-V^T)=0.60\pm0.01$ for NGC 5194, and $B^T=10.45\pm0.07$ mag and
$(B^T-V^T)=0.90\pm0.01$ for NGC 5195 \citep{dev91}.
At the adopted distance (without internal extinction correction), the absolute
magnitudes are $M_B^T=-20.81$ mag, $M_V^T=-21.38$ mag for NGC 5194,
and $M_B^T=-19.32$ mag and $M_V^T=-20.19$ mag for NGC 5195.

\section{Star Cluster Selection and Photometry}
\label{phot}

We detected and classified star clusters in M51 based on $HST$ ACS data and
made a catalog of the star clusters with $F555W<23$ mag. Detailed
information on star cluster detection, photometry, and classification was
described in \citet{hwa08} and only a brief summary is provided here.

We carried out the source detection in the background-subtracted $F555W$
band image of ACS data using SExtractor \citep{ber96}. A detection threshold
of $4$ $\sigma$ and minimum contiguous detected area of 5 pixels were used
for finding sources. The flux of the detected objects was measured using
SExtractor in dual mode for the $F435W$, $F555W$, and $F814W$ band
images. An aperture with $r=6$ pixels ($0.3\arcsec \simeq 12$ pc) was
adopted for the photometry. The instrumental magnitudes in the $F435W$,
$F555W$, and $F814W$ bands were calibrated using the photometric zero
points of the Vega magnitude system for the $HST$ ACS/WFC provided by
\citet{sir05}. We visually inspected candidate sources with $F555W<23$ mag
considering their radial profile, morphology, and environmental conditions. We
selected and classified star clusters into two different categories: Class 1 star
clusters that have a circular shape and no prominent nearby neighbors; and
Class 2 star clusters that have an elongated shape and/or irregular structure,
and/or multiple neighbors. Finally, we selected 2,224 Class 1 and 1,388 Class
2 star clusters with $F555W<23$ mag and their catalog was provided in
\citet{hwa08}.

We carried out an independent source detection and photometry in the
$F336W$ WFPC2 image data. Sources in WFPC2 image data were detected
using SExtractor with a detection threshold of $3$ $\sigma$ and minimum
contiguous detected area of 3 pixels ($0\arcsec.3$). The flux of detected
sources was measured using an aperture with $r=0\arcsec.3$ (6 pixels for PC
and 3 pixels for WF) in SExtractor. The instrumental magnitudes in $F336W$
were calibrated using the photometric zero points of the Vega magnitude
system for WFPC2 listed in the $HST$ Data Handbook for WFPC2. A correction
for the charge transfer efficiency (CTE) loss was made based on the
prescription given by \citet{dol00}.\footnote{ See
http://purcell.as.arizona.edu/wfpc2\_calib/ for further updated information.}
The coordinates of the detected sources were derived using $metric$ task in
STSDAS package of IRAF that corrects for the geometric distortion of each
chip in WFPC2. The derived coordinates of sources in WFPC2 data were
calibrated to the system of $HST$ ACS data.

We compared the $F336W$ band catalog with the catalog derived from the
$F435W$, $F555W$, and $F814W$ band data. There are about 1,400 Class 1
and about 1,000 Class 2 star clusters that have counterparts detected in the
$F336W$ band within a matching radius of $0\arcsec.4$. We checked the
spatial distribution of the sources that have no $F336W$ band counterpart
and the locations of WFPC2 observation fields to separate star clusters that
do not emit a significant flux in the $F336W$ band from those that were not
covered in the $F336W$ band observations. It is found that the number of
star clusters that were not observed in the $F336W$ band is about 300 for
Class 1 and about 80 for Class 2. Hereafter, we use $U$, $B$, $V$, and $I$
to denote $F336W$, $F435W$, $F555W$, and $F814W$, respectively.

\begin{figure}
 \plotone{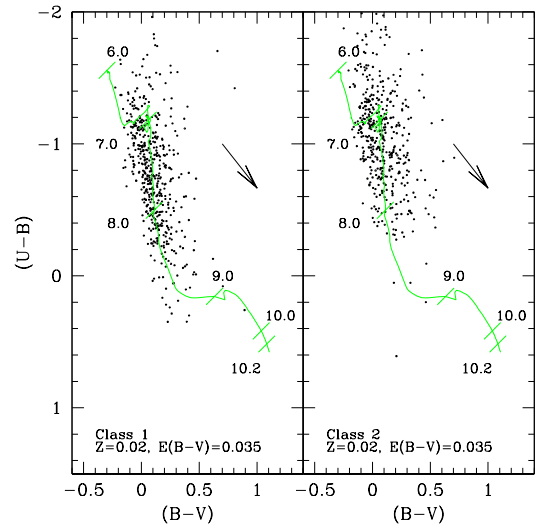} \caption{
 $(B-V)$ -- $(U-B)$ diagrams of the Class 1 star clusters (left panel)
 and Class 2 star clusters (right panel) with $V<23$ mag in M51.
 The solid line represents a theoretical evolutionary track with $Z=0.02$
 by \citet{bc03} shifted according to the foreground reddening $E(B-V)=0.035$.
 Numbers that run from 6.0 to 10.2 with tick marks on
 the track represent $\log$(age).
 The arrow in each panel shows the direction of reddening corresponding to $E(B-V)=0.3$.
 \label{ccd}}
\end{figure}

Figure \ref{ccd} shows the $(B-V)$ -- $(U-B)$ color-color diagram for the
star clusters (Class 1 and Class 2) in M51. We also display a theoretical
evolutionary track for the simple stellar population of \citet{bc03} with
metallicity $Z = 0.02$, shifted according to the foreground reddening $E(B-V)
= 0.035$. It is seen that most star clusters detected in $U$ band are blue
with $(U-B)<0.2$ and younger than 1 Gyr. There are only a few Class 2 star
clusters with $(U-B)>-0.4$ (older than about 100 Myr in age), while there are
still many Class 1 star clusters with $-0.4<(U-B)<0.0$. It is noted that the
$(U-B)$ color for the blue clusters varies from --1.8 to --0.2, while the
$(B-V)$ color varies only from 0.2 to --0.2. This shows the usefulness of $U$
band photometry in estimating the age of young star clusters.

We tested the completeness of star cluster detection in $V$ and $U$ band
images. Since the efficiency of star cluster detection can be affected by the
degree of source crowding, we selected two separate fields with different
source crowding for this test: Field A ($\alpha=13:30:03.2$,
$\delta=+47:13:32.6$) with higher crowding and Field B
($\alpha=13:30:05.8$, $\delta=+47:11:49.5$) with lower crowding. Each
field was set to a square with size of about $1\arcmin \times 1\arcmin$. The
selected two fields were covered in both $V$ and $U$ bands. Artificial star
clusters were generated using $mkobjects$ task of ARTDATA package in IRAF.
The point spread function (PSF) images required for artificial cluster generation
were prepared in ACS $V$ band data and in WFPC2 $U$ band data.

We generated 120 artificial clusters in one test field image over the magnitude
range of $18 \sim 23.5$ mag with $0.5$ mag step (10 artificial clusters per
each magnitude step). The FWHM and ellipticity of artificial clusters were
derived from the observed size and ellipticity distribution of M51 star clusters
in Table 1 of \citet{hwa08}. To do that, we modeled observed distributions
using a Gaussian centered at $3.25$ pixels with $\sigma = 2.0$ for FWHM and
another Gaussian centered at $0.15$ with $\sigma = 0.25$ for ellipticity.
Then, FWHM and ellipticity were randomly derived from these Gaussian
distributions and assigned to each artificial cluster. However, we selected only
artificial clusters with FWHM $> 2.0$ pixels to be consistent with the observed
data. We randomly distributed artificial clusters in a test field image and carried
out source detection and photometry in the same way as it was done for M51
star clusters. This routine of test was repeated 10 times in each test field and
the results of source detection and photometry were compared with the input
model parameters to calculate the completeness of our star cluster
photometry data.

\begin{figure}
\plotone{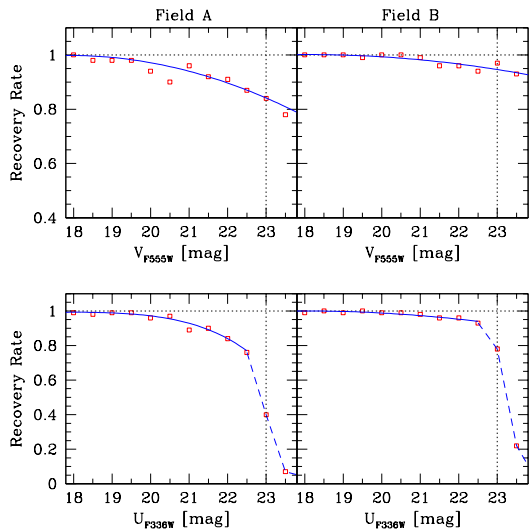}
\caption{Recovery rates of artificial star clusters as a function of
magnitudes in $V$ and $U$ band derived from Field A (with higher crowding) and B (with lower crowding).
Solid lines are polynomial fits to the data and dashed lines are straight lines connecting data points.
\label{artest}}
\end{figure}

Figure \ref{artest} shows the result of the completeness test for Field A (high
crowding) and Field B (low crowding). It can be seen that the recovery rate
(completeness) of artificial clusters with $V<23$ mag is higher than $80\%$
($>84\%$ for Field A and $>93\%$ for Field B). However, the recovery rate
of artificial clusters with $U=23$ mag drops to about $40\%$ in Field A and
about $80\%$ in Field B. This shows that the number of star clusters that we
may have failed to detect increases more significantly in the $U$ band than in
the $V$ band as the clusters get fainter. However, it is noted that the
recovery rate of star clusters is higher than $80\%$ when star clusters are
brighter than $U=22$ mag ($84\%$ for Field A and $96\%$ for Field B). This
value is consistent with the recovery rate of star clusters with $V=23$ mag,
the magnitude limit adopted for the star cluster survey in \citet{hwa08}.

\section{Age and Mass Estimation}
\label{sed}

We compared the photometric data of M51 star clusters with the model
spectral energy distribution (SED) to estimate the ages and masses of the star
clusters. Model SEDs were derived from the theoretical evolutionary synthesis
model by \citet{bc03}. For these models we adopted a Salpeter initial mass
function with a power slope of $x = -2.35$ \citep{sal55}, a lower mass cutoff
of 0.1 $M_{\odot}$ and an upper mass cutoff of 100 $M_{\odot}$. The
models span ages from 1 Myr to 15 Gyr for metallicity range of $Z=0.0001
\sim 0.05$. However, due to the age-metallicity-reddening degeneracy in the
integrated colors of star clusters, we assume the solar metallicity for our
working models to derive only age and reddening. Observations of the HII
regions in M51 shows that the current metallicity of
the gas in this galaxy is approximately solar (e.g., Diaz et al. 1991; Hill et al. 1997). 
It is also known that the influence of metallicity becomes more pronounced for
old star clusters with age of $> 1$ Gyr in their integrated colors. Because the
majority of star clusters in M51 is expected to be younger than 1 Gyr from
the study in \citet{bas05a} and \citet{hwa08}, the adoption of the solar
metallicity is considered appropriate.

We compared an SED model with a certain age to the observed data for each
star cluster, after reddening the model by $E(B-V)$ values between 0.0 and
0.6 in steps of 0.015, for about 2,400 star clusters with $UBVI$ photometry.
This range includes the reddening values derived from the previous studies for
the most star clusters in M51. \citet{lam02} showed that the reddening in the
bulge of NGC 5194 is typically $E(B-V) \approx 0.2$ and \citet{lee05} also
reported that most star clusters in M51 appear to have associated reddening
$E(B-V) <0.1$. For every combination of age and $E(B-V)$, we fitted the
theoretical SED models to the observed photometric data and calculated the
$\chi^2$. The fit with a minimum $\chi^2$ was adopted as the best fit for the
age and $E(B-V)$ combination.

\begin{figure}
 \plotone{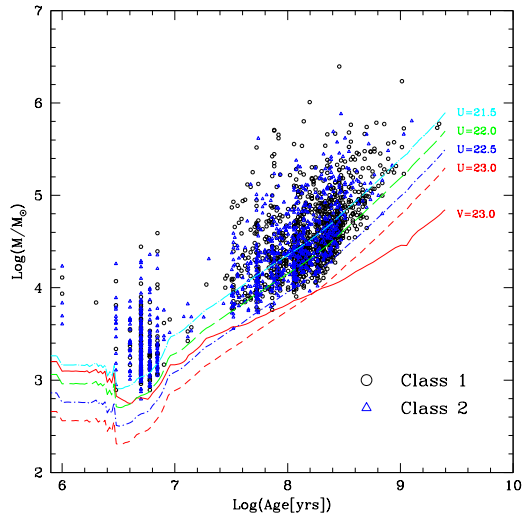}
 \caption{Mass vs. age of the star clusters in M51 (circles for Class 1 and
 triangles for Class 2).
 The solid line shows the magnitude
 limit of $V=23$ mag used to select star clusters in \citet{hwa08}.
 Other lines display magnitude limits in the $U$ band.
 \label{age2mass}}
\end{figure}

The number of finally accepted fits is 1,125 for Class 1 and 835 for Class 2
star clusters. Then we derived the masses for these star clusters by combining
the $M/L_V$ ratio of the best fit model with the measured $V$ band cluster
luminosity, extinction, and the distance to M51. Figure \ref{age2mass} shows
the age versus mass diagram for these star clusters. The solid curve indicates
the magnitude limit of $V=23$ mag adopted for star cluster selection in
\citet{hwa08}. The other lines show the expected age and mass of star
clusters with different $U$ band magnitude ranging from $23$ to $21.5$ mag.
This shows that star clusters with $1$ Gyr are detected only if they are more
massive than $10^{5.2} M_{\odot}$ (brighter than $U \approx 22$ mag) as
indicated in Figure \ref{age2mass}. Note that the completeness for $U=22$
mag band is $80\%$ as shown in Figure \ref{artest}.

Several features are noted in Figure \ref{age2mass}. First, there is a region
with few star clusters between $\log t =7$ and $7.5$ ($t$ represents an age
in units of years). This is considered to be an artifact that is caused by the
age-dating techniques that compare the integrated colors of star clusters with
the synthesis models. In this age range, the predicted colors of star clusters
span a very narrow range of colors, e.g., $\delta (U-B) \approx 0.16$ mag,
$\delta (B-V) \approx 0.00$ mag, and $\delta (V-I) \approx 0.22$ mag in the
models by \citet{bc03}. Therefore, even a small photometric error can result
in a fit to star cluster age younger than $ \log t=7$ or older than $ \log t
=7.5$. Second, there are two age regions of over abundance in star clusters:
one with $\log t \leq 7$ and the other with $7.5 < \log t < 9.0$. The
over-density at $\log t \leq 7$ is suspected to be, in part, due to the
under-density at $7.0 < \log t < 7.5$. Nonetheless, it is still a sign that shows
the existence of many young star clusters, which is consistent with the results
of several studies on HII regions (e.g., Scoville et al. 2001; Lee, J. H. et al.
2009 in preparation). The other over-density at $7.5 < \log t < 9.0$ indicates
that many star clusters were actually formed in this epoch. Third, there is an
apparent trend that the mass of the most massive clusters increases with age.
This is known as a `size-of-sample effect' noted by \citet{hun03}. However,
it seems that the upper limit of this relation is composed of multiple
components. We discuss this further in Section \ref{disc2}.

Similar features in the age and mass distribution were also reported in several
other studies of M51 star clusters. For example, the under-density of star
clusters in $7.0 < \log t < 7.5$ can be noted in the results given by
\citet{bas05a} (their Figure 10) and \citet{gie05} (their Figure 1) in
combination with the strong over-density of star clusters with $\log t < 7$.
This is consistent with the result shown in Figure \ref{age2mass}. However,
\citet{sch09} show rather different age and mass distributions in their study
of M51 star clusters using the same data that are used in this study. Most star
clusters in the sample of \citet{sch09} are found to be younger than $\log t =
7.5$ and there is no evident feature of under-density in $7.0 < \log t < 7.5$,
which is contrary to the results of other studies \citep{bas05a, gie05} as well
as this study. One of the possible reasons for this difference may be different
criteria adopted for star cluster selection, which may lead to a different sample
of star clusters used for the analysis.

Another noteworthy point shown in \citet{sch09} is that different population
synthesis models can introduce a systematic change to the derived age and
mass distributions of star clusters. However, the result of their study shown in
their Figure 4 reveals that the age distribution of star clusters with $\log t
\geq 8$ is found to be more-or-less consistent with one another even in the
case that different isochrones (Padova or Geneva) and different metallicities
are adopted for models. Considering these inconsistencies and limits caused by
the different analysis methods adopted, we focus on the age and mass
distributions of star clusters with $\log t > 7.5$.

\section{Results}
\label{result}

\subsection{Star Cluster Age Distribution}
\label{agedist}

\begin{figure}
 \plotone{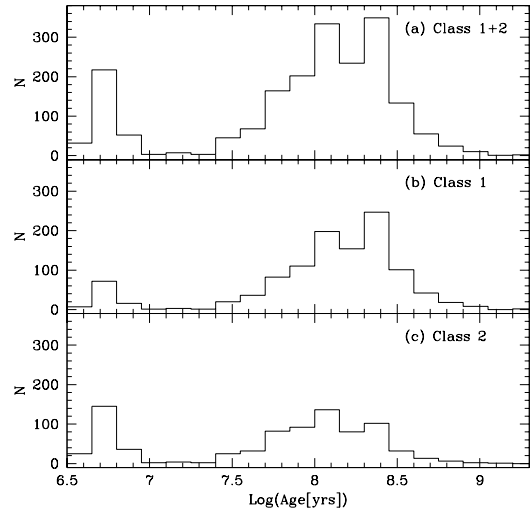}
 \caption{
 Age distributions of the star clusters in M51:(a) all (Class 1 and 2) clusters, (b) Class 1 clusters,
 and (c) Class 2 clusters.
 Three peaks at $\log t \approx 6.7$, 8.0, and 8.4 are noted.
 \label{agehist}}
\end{figure}

Figure \ref{agehist} shows the age distributions of all star clusters (including
both Class 1 and Class 2), Class 1 star clusters, and Class 2 star clusters. It is
clear that the age distributions display two over-dense epochs: one with $\log
t<7$ and another with $7.5< \log t<9.0$. The over-densities of star clusters
in these two periods are evident in the age distribution of both Class 1 and
Class 2 star clusters, as shown in Figures \ref{agehist}(b) and (c). However,
about 1000 Class 1 star clusters are found in the epoch of $7.5< \log t<9.0$,
while only about 100 Class 1 star clusters are younger than $\log t =7$. On
the other hand, for Class 2 star clusters, the number of Class 2 star clusters
younger than $\log t=7$ is about 230, about one third of the number of star
clusters with $7.5<\log t<9.0$.

One interesting point is that there are distinct features in the age range of
$7.5 < \log t< 9.0$ in Figure \ref{agehist}(a). The most pronounced one is a
broad component over $\log t \approx 7.8$ -- $8.5$ that displays two peaks
at $\log t \approx 8.0$ and $\log t \approx 8.4$. The same broad component
is also present in the distributions of Class 1 and Class 2 star clusters shown
in Figures \ref{agehist}(b) and (c), respectively. However, for Class 1 star
clusters, the peak at $\log t \approx 8.4$ is more enhanced than that at $\log
t \approx 8.0$. On the other hand, the peak at $\log t \approx 8.0$ is
relatively more significant in case of Class 2 star clusters. Other than these,
the age distributions of Class 1 and Class 2 clusters are consistent with each
other. Therefore, we consider the age distribution of all star clusters including
Class 1 and Class 2 clusters for the further analysis.

\begin{figure}
 \plotone{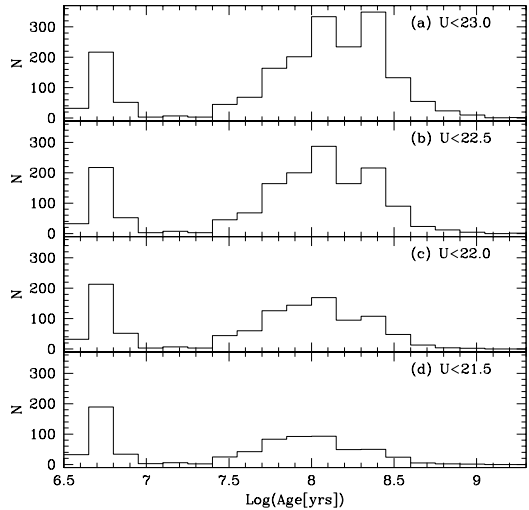}
 \caption{Age distributions of the star clusters in M51 for different $U$ band magnitude
  limits: (a) $U<23.0$ mag, (b) $U<22.5$ mag, (c) $U<22.0$ mag, and (d) $U<21.5$ mag.
  Note that three peaks at $\log t \approx 8.0$ and $8.4$ are visible clearly in all panels,
  and the oldest peak  at $\log t \approx 8.4$ is weak for the brightest limit $U<21.5$ mag.
  }
 \label{agedist_umag}
\end{figure}

We investigated the effect of incompleteness to the age distribution of star
clusters. As shown in Figure \ref{age2mass}, the incompleteness with
different $U$ band magnitudes may bias the interpretation of age distribution.
Therefore, we derived the age distribution of star clusters changing the limiting
$U$ band magnitude starting from $U=23.0$ to $21.5$ mag, and showed the
results in Figure \ref{agedist_umag}. Although the amplitude of histograms in
$7.5<\log t<9.0$ is gradually decreasing as the $U$ band magnitude limit gets
brighter from $U<23$ to $U<22$ mag, the overall features in the age
distribution remain consistent. Even when the $U$ band magnitude limit is
raised to $U=21.5$ mag, where the completeness is estimated to be about
$90\%$ (see Figure \ref{artest}), some hint of peaks at $\log t \approx 8.0$
and $\log t \approx 8.4$ can be noticed, as shown in Figure
\ref{agedist_umag}(d). This indicates that the two peaks at $\log t \approx
8.0$ and $\log t \approx 8.4$ in the age distribution of star clusters shown in
Figure \ref{agehist} are reliable against any incompleteness problem expected
from the photometry.

\subsection{Star Cluster Mass Function}
\label{masdist}

\begin{figure}
 \plotone{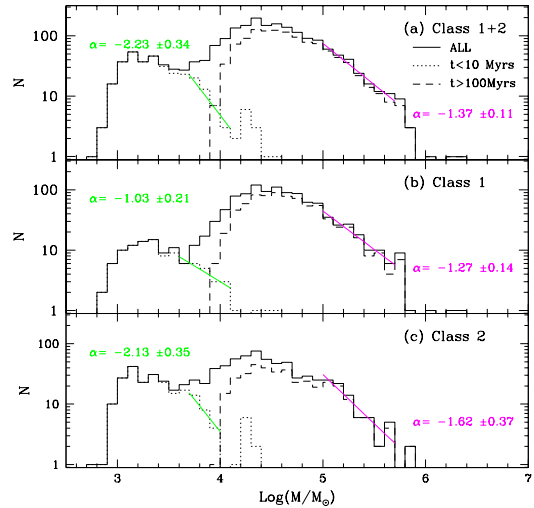}
 \caption{
 Mass functions of the star clusters in M51: (a) all (Class 1 and 2) clusters,
 (b) Class 1 clusters, and (c) Class 2 clusters (solid line histograms).
 Dotted line and dashed line histograms represent very young star clusters with $t<10$ Myr,
 and the intermediate-age star clusters with $t>100$ Myr, respectively.
 Solid lines represent power law fits (linear fits in the logarithmic scales) for given mass ranges.
 \label{mashist}}
\end{figure}

Figure \ref{mashist} shows the mass functions of all star clusters (including
both Class 1 and Class 2), Class 1 star clusters, and Class 2 star clusters in
M51. It is found that the mass of star clusters in M51 ranges from $10^3$ to
$10^6 M_{\odot}$ for both Class 1 and Class 2 star clusters and that the
overall mass function appears to be separated into two components by a dip
at $\approx 10^{3.7} M_{\odot}$. This dip, separating low mass and high
mass clusters, is caused by an artifact of the SED fitting method over $\log t
= 7.0$ -- $7.5$ as described in Section~\ref{sed} and the star cluster
detection limit of $V<23$ mag adopted in this study, as shown in Figure
\ref{age2mass}. The star clusters with different masses also represent star
clusters with different ages, as shown in each panel of Figure \ref{mashist}:
the low mass star clusters are younger than $10$ Myr old (dotted lines) and
the high mass star clusters are mostly older than $100$ Myr old (dashed
lines).

The mass functions of young ($<10$ Myr) and intermediate-age/old ($>100$
Myr) star clusters can be represented with a power law over the range of
$3.6<\log (M/M_{\odot})<4.1$ for young clusters and $5.0<\log
(M/M_{\odot})<5.7$ for intermediate-age/old clusters, regardless of Class 1
or Class 2. It is noted that the upper limit for intermediate-age/old clusters is
marked by a weak bump at $\log (M/M_{\odot}) \sim 5.8$, as shown in each
panel of Figure \ref{mashist}. We derived a power law index $\alpha$ of the
cluster mass function as defined in $N dM \propto M^{\alpha} dM$. It is found
from the sum of Class 1 and Class 2 clusters that the power law index
$\alpha$ for old star clusters with $5.0<{\rm Log}(M/M_{\odot})<5.7$ is
$-1.37 \pm 0.11$. For star clusters with $\log (M/M_{\odot})<4.5$, where
the mass function of massive star clusters starts to flatten out, the index
$\alpha$, derived using star clusters younger than $10$ Myr, is $\alpha=-2.23
\pm 0.34$. This index $\alpha$ of young star clusters is in agreement with the
cluster mass function index $\alpha = -2.1 \pm 0.3$ reported by
\citet{bik03} but it is slightly lower than $\alpha = -1.70 \pm 0.08$ given by
\citet{gie06b} for M51 star clusters. It is, however, in agreement with the
cluster initial mass function (CIMF) index $\alpha \approx -2.0$ of star
clusters in late type galaxies \citep{zha99,deG03,gie06b}.

The overall shape of the mass function of Class 1 and Class 2 star clusters is
similar to each other. However, the mass function of Class 2 clusters appears
to be slightly steeper than that of Class 1 clusters. That is, the power law
index $\alpha$ is $-2.13 \pm 0.35$ for young ($<10$ Myr) and $-1.62 \pm
0.37$ for old ($>100$ Myr) Class 2 clusters, while $\alpha = -1.03 \pm0.21$
for young and $-1.27 \pm 0.14$ for old Class 1 clusters. Although the indices
for old clusters may be in agreement with each other considering a relatively
large error associated with Class 2 clusters, the difference in the mass
function indices for young Class 1 and Class 2 clusters is rather significant. As
explained in \citet{hwa08} and in Section \ref{phot}, one big difference
between Class 1 and Class 2 clusters is their morphologies: Class 1 clusters
are in circular shape, while Class 2 clusters have elongated or irregular
structures. This suggests that many low mass clusters are usually formed or
embedded in elongated and irregular structures.

To test the effect of incompleteness on the star cluster mass function, we
have investigated how different limiting magnitudes in $U$ and $V$ band
affect the overall shape and the slope of the mass function of star clusters.
We constructed the cluster mass function using star clusters selected with
varying limiting magnitude conditions in $U$ and $V$ bands ranging from $23$
to $21.5$ mag, respectively, and derived a power law index $\alpha$ of the
mass function over the same mass range as shown in Figure \ref{mashist}.
The resulting index $\alpha$ for young clusters ($\log t<7$) turns out to be
the same under the varying limiting magnitudes from $U<23$ to $U<21.5$
mag and from $V<23$ to $V<22$ mag, respectively. Even when the
constraint is set to $V<21.5$ mag, the mass function index is $\alpha = -1.88
\pm 0.57$, which is still in agreement with $\alpha = -2.23 \pm 0.34$ for
young clusters shown in Figure \ref{mashist}(a). For old clusters ($\log
t>8$), it is also found that the mass function index is $\alpha \approx -1.30
\pm 0.20$ for all magnitude constraints in $U$ and $V$ bands. This indicates
that the power law indices of the young and old cluster mass functions derived
in this study are affected little by the incompleteness problem. This is because
the power law fits were done using the mass limits set sufficiently high enough
to avoid severe incompleteness problem.

\subsection{Evolution of the Star Cluster Mass Function}
\label{massevol}

\begin{figure}
 \plotone{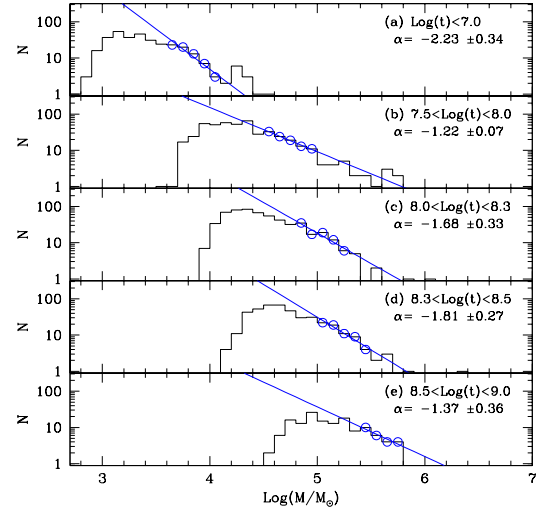}
 \caption{
 Mass functions of the star clusters in M51 for different age ranges (including both Class 1 and Class 2):
 (a) $\log t < 7.0$,  (b) $7.5 < \log t < 8.0$,  (c) $8.0< \log t < 8.3$,  (d) $8.3< \log t < 8.5$,  and
 (e) $8.5 < \log t < 9.0$.
 In each panel, the power law index of mass function is derived by fitting data points
 marked by circles and the solid line represents the best fit.
Solid lines represent the power law fits to the data (marked by circles) for given mass ranges,
and the resulting power law index for each age range is given in each panel (including both Class 1 and Class 2).
 \label{masfunc}}
\end{figure}

The variation of the slope of mass function depending on star cluster age
shown in Figure \ref{mashist} suggests that the index of cluster mass
function may be a function of time. The cluster mass function for given time is
determined by the initial cluster mass function, the cluster disruption, and the
fading below the detection limit,  as well as the variable cluster formation rate.
We investigated the mass function of star clusters that formed in different
epochs and derived its power law index for each epoch using the combined set
of Class 1 and 2 star clusters, and the result is shown in Figure
\ref{masfunc}. Age ranges were chosen to separate star clusters roughly
according to different features in the age distribution, as shown in Figure
\ref{agehist}.

Figures \ref{masfunc}(a) through \ref{masfunc}(e) show that the upper part
in the mass function of star clusters in each epoch can be represented
approximately by a single power law and that the slopes of mass function
change depending on the epochs. The derived power law indices of mass
function are $\alpha=-2.23 \pm 0.34$ for $\log t<7.0$ (panel a),
$\alpha=-1.22 \pm 0.07$ for $7.5<\log t <8.0$ (panel b), $\alpha = -1.68
\pm 0.33$ for $8.0<\log t<8.3$ (panel c), $\alpha = -1.81 \pm 0.27$ for
$8.3<\log t<8.5$ (panel d), and $\alpha = -1.37 \pm 0.36$ for $8.5<\log
t<9.0$ (panel e). These results are consistent with those in Figure
\ref{mashist} in the sense that the mass function of star clusters younger
than $\log t=7$ is steeper than that of older star clusters. However, one
interesting point in Figure \ref{masfunc} is that the mass function of star
clusters with $7.5< \log t <8.5$ gets steeper as ages increase. Especially, for
star clusters with $8.3<\log t <8.5$ shown in Figure \ref{masfunc}(d), the
power law index $\alpha = -1.81 \pm 0.27 $ is roughly in agreement with that
of young star clusters shown in Figure \ref{masfunc}(a). It is also noted that
the relatively shallow mass function of star clusters with $\log t>8.5$
($\alpha = -1.37 \pm 0.36$) is consistent with that of star clusters with $\log
t>8.0$ ($\alpha = -1.37 \pm 0.11$) shown in Figure \ref{mashist}.

Since the cluster mass function index changes systematically depending on the
epoch of star cluster formation, we investigated the power law index of cluster
mass function using star clusters with $\log t = 7.5$ -- $8.6$ when the
condition of star cluster formation is expected to change drastically. The width
of the logarithmic age bin was set to $\delta \log t=0.20$ and $0.15$ to keep
the number of star clusters more than 100 in each logarithmic bin, whenever it
is possible. The maximum number of clusters in a single bin is about 420 and
the minimum is about 70. We verified that the resulting cluster mass function
in each logarithmic age bin can be always represented by a power law function
and derived the index of mass function.

\begin{figure}
 \plotone{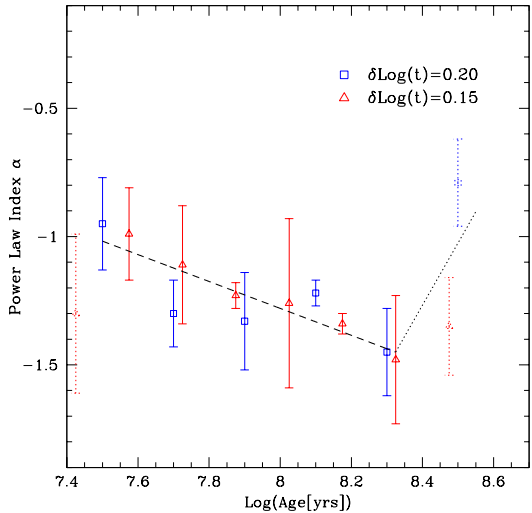}
 \caption{
 Mass function power law index vs. age of the star clusters in M51 (including both Class 1 and Class 2).
 The mass function index was derived for two logarithmic age bins,
 $\delta \log t = 0.15$ (triangles) and 0.20 (squares).
 Note that the mass function index shows a minimum value (the steepest) at $\log t \approx 8.3$.
 Approximate linear fits are drawn for $(t)<8.3$ (dashed line) and $\log t>8.3$ (dotted line).
 \label{slope}}
\end{figure}

Figure \ref{slope} displays the power law indices of mass functions derived
from logarithmic age bins with $\delta \log t=0.20$ (squares) and $0.15$
(triangles). If the power law index was derived using more than 100 star
clusters in each age bin, then the corresponding symbol is plotted in solid line.
Otherwise, the symbol is marked in dotted line. Figure \ref{slope} shows that,
as we move to the higher age domain starting from $\log t \approx 7.5$, the
star cluster mass function gets steeper, reaching $\alpha \approx -1.50$ at
$\log t \approx 8.3$, as shown in the dashed line derived from a simple linear
fit. Then, the trend is reversed and the mass function gets shallower again for
$\log t>8.4$. However, the number of star clusters used to derive the index
at $\log t \approx 8.5$ is smaller than $100$ (about $70$) and the scatter
between indices marked by a square and a triangle in dotted lines is large. This
suggests significant uncertainties due to the limited number of massive star
clusters available in this age domain (see Figure \ref{age2mass}).

Interestingly, the epoch of $\log t \approx 8.3$, when the cluster mass
function index seems to attain its minimum value, almost coincides with the
age distribution peak at $\log t \approx 8.4$ as shown in Figure \ref{agehist}.
This period is also coincident with the epoch of dynamical encounter of the
two galaxies NGC 5194 and NGC 5195 expected by theoretical models
\citep{too72, sal00a}. Therefore, the minimum value of cluster mass function
index around this period may have some implications related with the
increased number of star clusters and the dynamical interactions of the two
galaxies in the M51 system. However, if the cluster mass function was
affected and the number of star clusters increased during the course of
dynamical interactions of galaxies, then there should be some impacts on the
star cluster formation rate, which we focus on in the following section.

\subsection{Star Cluster Formation Rate}

The cluster formation rate is a good indicator of the star formation activity.
However, one of the difficulties of investigating the star cluster formation
history is that more star clusters disappear due to the fading and/or the
disruption as well as the observation incompleteness as we analyze older star
cluster populations. Another difficulty is that, as shown in Figures
\ref{masfunc} and \ref{slope}, the star cluster mass function does not
appear to be constant but to evolve depending on the formation epochs. This
suggests that the number of faint star clusters that we failed to detect may be
different in different age bins, which makes the estimation of the number of
the lost star clusters even more complicated.

Although the slope of the star cluster mass function depends on the epoch,
the mass function itself can always be represented by a power law function, as
shown in Figure \ref{masfunc}. Therefore, based on the assumption that
every star cluster mass function has a power law form, we constructed a
`simple model' mass function of star clusters by extrapolating the power law
fitted mass function over $\log t \approx 7.5$ -- $8.6$ period. This
extrapolation method allows to compensate the number of star clusters lost
due to the detection limit and the incompleteness problem of the observed
data. However, this method does \emph{not} correct for star cluster
disruption. For the model mass function, we set the lower mass limit to
$10^3$ $M_{\odot}$ and the upper mass limit to $10^{6.5}$
$M_{\odot}$, since there is no star cluster with more massive than
$10^{6.5}$ $M_{\odot}$ in our data, as shown in Figure \ref{mashist}.

Using these simple model and observed mass functions, we calculated the
number of star clusters per each Myr, i.e., the star cluster formation rate, in
each logarithmic age bin over $7.5<\log t<8.6$. For this investigation, two
mass limited samples of star clusters were chosen by selecting star clusters
with $\log (M/M_{\odot})>5.0$ and $4.6<\log  (M/M_{\odot})<5.0$. These
criteria were defined to select star clusters located above the detection limit of
$U=22$ mag (see Figure \ref{age2mass}) and the completeness of about
$80\%$ (see Figure \ref{artest}) at $\log t \lesssim 8.8$  for $\log
(M/M_{\odot})>5.0$ and at $\log t \lesssim 8.5$ for $4.6< \log
(M/M_{\odot})<5.0$. The definition of mass limited cluster samples also
helps to avoid any possible artificial contribution from the low mass part of the
model mass function, and enables to test whether any discrepancy in cluster
formation rate exists between different mass ranges.

\begin{figure}
 \plotone{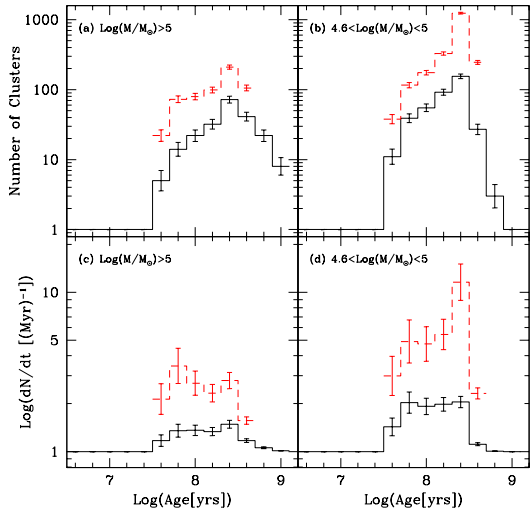}
 \caption{
 Age distributions (a and b) and cluster formation rates as a function of age (c and d)
 for two mass ranges: $\log (M/M_{\odot})>5.0$ (a and c)
 and $4.6<\log (M/M_{\odot})<5.0$ (b and d).
 Solid line histograms represent the observed data
 and dashed line histograms represent the simple model data.
 Note that there is seen a peak at $\log t \approx 8.4$
 in the age distributions for both mass ranges and
 in the cluster formation rate derived from the model data for the lower mass range.
 \label{cfr2}}
\end{figure}

Figure \ref{cfr2} displays the age distribution and cluster formation rate
derived from star clusters with $\log (M/M_{\odot})>5.0$ (panels a and c)
and $4.6<\log (M/M_{\odot})<5.0$ (panels b and d). Solid lines are
from the observed data, while dashed lines are from the model mass function.
The observed data of star clusters with $\log (M/M_{\odot})>5.0$ reveal a
peak in the age distribution and an abrupt increase of the cluster formation
rate at $\log t \approx 8.4$, as shown in Figures \ref{cfr2}(a) and (c).
Then, the data exhibit a gradual decrease in the age distribution and a rough
plateau in the cluster formation rate in $7.8<\log t<8.4$. Very similar
features are seen in the age distribution and cluster formation rate of star
clusters with $4.6<\log (M/M_{\odot})<5.0$ in Figures \ref{cfr2}(b)
and (d). The peak at $\log t \approx 8.4$ shown in the age distribution and
the cluster formation rate is consistent with the peak in the age distribution found
around $\log t = 8.4$ as shown in Figure \ref{agehist}.

The data from model mass function also display a peak at $\log t \approx
8.4$ in the age distribution shown by dashed lines in Figures \ref{cfr2}(a) and
(b).  However, for star clusters with $\log (M/M_{\odot})>5.0$, one slightly
different feature is a bump-like peak at $\log t \approx 7.8$ in the age
distribution in Figure \ref{cfr2}(a). Although this feature does not appear
significant, it seems to translate into another peak of the cluster formation
rate at $\log t \approx 7.8$ apart from the one at $\log t \approx 8.4$ in
Figure \ref{cfr2}(c). For star clusters with $4.6<\log (M/M_{\odot})<5.0$,
no such a bump-like feature is seen in the age distribution and no cluster
formation rate peak exists at $\log t \approx 7.8$, as shown in Figures
\ref{cfr2}(b) and (d). The model mass function data reproduce main features
exhibited by the observed data, including a peak at $\log t \approx 8.4$ and a
gradual decrease in the age distribution, and a plateau in the cluster formation
rate in $7.8<\log t<8.4$.

Figure \ref{cfr2} shows that star cluster formation rate was raised at $\log t
\approx 8.4$ and was maintained in that heightened state for somewhat
extended period until $\log t \approx 7.8$. This is true for star clusters with
both $\log (M/M_{\odot})>5.0$ and $4.6< \log(M/M_{\odot})<5.0$.
However, it is also probable that massive star clusters with $\log
(M/M_{\odot})>5.0$ have experienced more than one burst of cluster
formation: one at $\log t \approx 8.4$ and another at $\log t \approx 7.8$.
On the other hand, according to the model mass function, star clusters with
$4.6< \log (M/M_{\odot})<5.0$ exhibit one strong peak of the cluster
formation rate at $\log t \approx 8.4$.

\section{Discussion}
\label{discuss}

\subsection{Star Cluster Formation and Dynamical Interactions of
Galaxies}

The age distribution shown in Figure \ref{agehist} displays two peaks at $\log
t \approx 8.0$ and $\approx 8.4$. Interestingly, the epochs of these two
peaks coincide with the dynamical interaction epochs predicted by the multiple
encounter model of M51 system \citep{sal00a}. However, if the size of
logarithmic age bin is enlarged from the current value $\delta \log t = 0.15$
to $\delta \log t = 0.2$, then those two age peaks merge into one broad
component ranging over  $\log t = 8.0$ -- $8.4$. In this case, it is not
straightforward to confirm the existence of independent peaks in the star
cluster age distribution and the possible correlation between those peaks and
the individual dynamical interactions, even though it is still clear that a large
population of star clusters were formed during this period of $\log t = 8.0
\sim 8.4$. The uncertainty is mostly due to the typical error in age estimation
based on the broad band SED fit technique, about $20\%$ \citep{and04,
deG03}.

Figure \ref{cfr2} shows that the star cluster formation rate increased over
the same period of $\log t = 8.0 \sim 8.4$. Moreover, more than one event
of significant cluster formation rate increase are apparent at $\log t \approx
8.4$ and $\log t \approx 7.8$, depending on the mass of star clusters.
However, the cluster formation peak at $\log t \approx 7.8$ is slightly
apparent only from the model data for massive star clusters with
Log$(M/M_{\odot})>5$. It may be possible that the peak at $\log t \approx
7.8$ was made by some stochastic effect and that it may not be a real
feature because the peak is not so strong and involves a relatively large error.
However, this recent peak in the cluster formation rate is noted to be nearly
coeval with the peak in the age distribution at $\log t \approx 8.0$, as shown
in Figure \ref{agehist}.

Except for the SED fit error, the star cluster disruption and the evolution of
luminosity and mass function can affect the star cluster age distribution and
cluster formation rate. Although many uncertainties are related with these, it
is relatively well known from the theoretical and observational studies
\citep{fal06, lar09} that the luminosity function and mass function of young
star clusters follow, in most cases, power law functions with a single
parameter $\alpha$ that defines its slope. \citet{hwa08} showed that the
luminosity function of M51 star clusters is fitted by a single power law with
$\alpha = -2.59 \pm 0.03$, which is in good agreement with the results in
other studies of M51 star clusters, $\alpha = -2.5 \pm 0.1$ by \citet{gie06b}
and $\alpha = -2.53 \pm 0.06$ by \citet{has08}.

Figure \ref{masfunc} shows that the mass function of star clusters with
different formation epochs can be also described with a single power law,
although the mass function index seems to change depending on the epochs.
Even when we select star clusters from a single logarithmic age bin to
construct a mass function, the resulting mass function is found to follow a
single power law over a period of $\log t \approx 7.5 \sim 8.6$. However,
Figure \ref{slope} shows that the mass function appears to evolve depending
on the cluster formation epoch. That is, the mass function gets steeper as star
clusters get older until $\log t \approx 8.3$ when the index reaches the
minimum value $\alpha \approx -1.50$, making the steepest mass function.
Then, the mass function appears to overturn to become a shallower function
as star clusters get older than $\log t \approx 8.3$.

The change in the mass function slope may be explained as a result of the
evolution of star clusters. Especially, the cluster mass function can be
shallower due to the selective disruption of low mass star clusters. This is one
of the reasons why the mass function of star clusters older than $100$ Myr
($\alpha = -1.37 \pm 0.11$) is flatter than that of star clusters younger than
$10$ Myr ($\alpha = -2.23 \pm 0.34$) as noted in Section \ref{masdist}.
However, Figure \ref{slope} shows that the mass function gets steeper
reaching the power law index $\alpha \approx -1.50$ at $\log t \approx 8.3$.
If we assume that the star cluster disruption and evolutionary fading should
apply to all star clusters consistently regardless of their formation epochs, the
change in the mass function slope suggests that the number of star clusters
increased significantly around the epoch of $\log t = 8.3$.

What our results show is that many star clusters are formed around $\log t =
8.4$ ($250$ Myr ago) and this period is roughly in agreement with the first
encounter period of NGC 5194 and NGC 5195 predicted by theoretical models
\citep{too72,sal00a}. This may suggest that the dynamical interaction
between two galaxies in the M51 system induced star cluster formation,
implying a strong correlation between the active star cluster formation and the
dynamical interactions experienced by the host galaxies. This is also in
agreement with the result of \citet{lee05} that the number of star clusters
with $100$ -- $400$ Myr is significantly larger in M51 than in other late type
galaxy M101, suggesting a correlation between the increased number of star
clusters and the dynamical interaction(s) at the similar epoch.

There is a hint, although weak, for another increase of cluster formation rates
around $\log t =7.8$, e.g., as shown in Figure \ref{cfr2}(c). In a study on the
cluster formation in M51, \citet{bas05a} reported the existence of relatively
recent star cluster bursts in M51: one at $\approx 6$ Myr ($\log t \approx
6.8$) and another at $\approx 60$ Myr ($\log  \approx 7.8$). The latter
burst coincides with the probable increase of star cluster formation at $\log t
\approx 7.8$. The former star cluster burst is reproduced as a peak at $\log t
\approx 6.7$ in the star cluster age distribution shown in Figure \ref{agehist}.
No report about the increase of star cluster formation in $\log t>8.0$ was
given in \citet{bas05a}. However, it is noted that a weak hint is apparent in
their Figure 12 that implies a slight increase of cluster formation rate at $\log
t \approx 8.5$ for star clusters with $\log (M/M_{\odot})>4.7$.

\subsection{Star Cluster Disruption and Age Distribution}
\label{disc2}

Among many studies in the literature, \citet{lam08} described very concisely
two major models that are used for studies of star cluster age distribution,
i.e., the `Baltimore model' \citep{fal05} and the `Utrecht model'
\citep{bou03}. The Baltimore model is characterized by the mass independent
disruption of about 90\% star clusters in each age dex up to approximately 1
Gyr. On the other hand, the Utrecht model includes the evolutionary fading
and the mass dependent dynamical cluster dissolution by tidal effects as well
as the mass independent dissolution of star clusters in their very early stages,
which they termed as `infant mortality'. These two models predict different
outcomes in the distribution of maximum cluster mass in each age dex
\citep{gie08} and in the cluster age distribution in a $\log (dN/dt)$ versus
$\log (t)$ plane \citep{lam08}. We use our age and mass data of M51 star
clusters to test these two models.

\begin{figure}
 \plotone{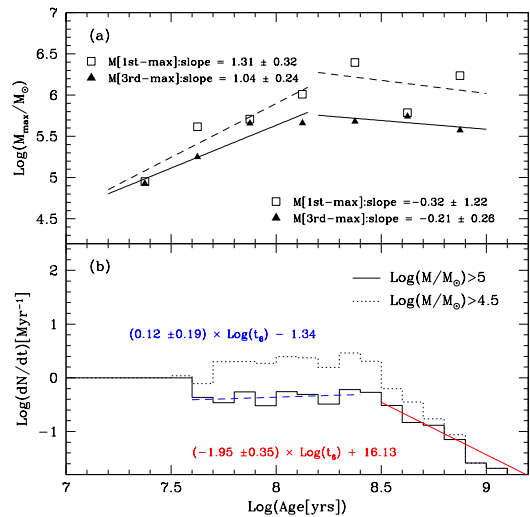}
 \caption{
(a) The 1st (squares) and 3rd (triangles) maximum mass vs. age of the star clusters in M51
(including both Class 1 and Class 2).
 Note that there is a break at $\log t \approx 8.2$.
Dashed lines and solid lines represent linear fits to the data
for the 1st and 3rd maximum masses, respectively.
(b) Cluster formation rate vs. age of the star clusters in M51 (including both Class 1 and Class 2)
for two mass ranges: $\log (M/M_{\odot})>5)$ (solid line histogram) and
$\log (M/M_{\odot}) >4.5)$ (dotted line histogram).
 Linear fits to flat and steep parts are plotted in dashed and solid lines, respectively.
 \label{dndt}}
\end{figure}

One way to derive the cluster disruption rate is to use the relation between
mass and age of star clusters and a `size of sample' effect that more massive
star clusters are found in older logarithmic age bins, as shown in Figure
\ref{age2mass}. This effect comes from the assumption that when star
clusters are randomly selected from a physically determined mass function,
the larger samples are likely to have the more massive star clusters.
\citet{hun03} described how to derive the CIMF index $\alpha$, independent
of cluster disruption and evolutionary fading rates from the slope defined using
the maximum mass and the age of star clusters, assuming a constant cluster
formation. \citet{gie08} expanded this by showing that this slope is also an
indicator for the mass independent disruption rate of star clusters when we
have information about the CIMF. The Baltimore model expects about 90\%
star clusters in each logarithmic age bin to disrupt regardless of their mass,
which would result in a flat distribution as shown in \citet{gie08}.

Figure \ref{age2mass} shows that, although it is hard to define due to large
scatters, the upper limit of mass and age distribution of M51 star clusters
appears to be composed of roughly two linear components: $7.0< \log
t<8.2$ and and $\log t>8.2$. Figure \ref{dndt}(a) displays this distribution of
the maximum mass (open square) and the 3rd maximum mass (filled triangle)
of M51 star clusters in each logarithmic age bin with a step of $\delta \log t =
0.25$. A linear square fit to $\log t<8.2$ returns a slope of $1.31 \pm 0.32$
for the maximum mass and $1.04 \pm 0.24$ for the 3rd maximum mass,
while another fit to $\log t>8.2$ returns $-0.32 \pm 1.22$ for the maximum
mass and $-0.21 \pm 0.26$ for the 3rd maximum mass. If we fit this
distribution with a single line over the range of $7.0<\log t<9.5$, the slope
turns out to be $0.72 \pm 0.23$. This value is steeper than $0.4$ reported by
\citet{lee05}, $0.26 \pm 0.09$ by \citet{gie06b}, and $0.23$ by
\citet{gie08} for M51 star clusters.

This shows that the maximum mass distribution for young star clusters with
$\log t<8.2$ is steep, while that for old star clusters with $\log t>8.2$ is
more or less flat. The slope of the steep part is even slightly steeper than
$0.74$ for LMC and $0.69$ for SMC reported by \citet{hun03}, although the
slope of about $0.72$ derived over $7.0< \log t <9.5$ is in good agreement
with these. However, the steep slope derived from star clusters with $\log t
<8.2$ is consistent with those derived from LMC ($0.96$) and SMC ($0.93$)
clusters with Log$(t)<8.0$ by \citet{gie08}. On the other hand, for the
Antennae galaxy, \citet{gie08} show that the slope is $0.01$ for clusters with
Log$(t)<8.0$, which is consistent with the Baltimore model. These results
suggest that the Baltimore model, at least, does not apply to M51 star
clusters, considering that the flat distribution of maximum mass in $\log t >
8.2$ may be due to the physical upper limit of M51 star cluster mass
\citep{gie06b}.

Figure \ref{dndt}(b) displays the age distribution of the mass limited star
clusters in a $\log dN/dt$ versus $\log t$ plane. It clearly shows a bend at
$\log t \approx 8.4$ and the existence of a plateau over $\log t \approx 7.6$
-- $8.4$, which is a characteristic of the Utrecht model. On the contrary, the
Baltimore model expects a consistently decreasing relation as shown by
\citet{fal05} in their Figure 2. Even when we change the mass limit, the result
is still consistent for $\log (M/M_{\odot})>5.0$ (solid line) and for $\log
(M/M_{\odot})>4.5$ (dotted line). Applying a simple linear fit returns a slope
of $0.12 \pm 0.19$ for the plateau (dashed line) and $-1.95 \pm 0.35$ for
the steep part at $\log t > 8.4$ (solid line). This slope $-1.95 \pm 0.35$ of
the steep part is in agreement with the value ($\approx -1.7$) expected by
the Utrecht model considering the error \citep{lam08}. If we apply the
analysis of \citet{bou03} relating this slope $A$ to the CIMF slope $\alpha$,
we can derive the dynamical disruption parameter $\gamma$ from the
equation $A = (1+\alpha)/\gamma$ (for $\alpha<0$). If we assume the CIMF
with $\alpha = -2.0$, then the disruption parameter is $\gamma = 0.53 \pm
0.10$. This is in good agreement with the result $0.57 \pm 0.10$ reported by
\citet{lam05} for M51. Therefore, the age distribution of M51 star clusters
can be described better by the Utrecht model than by the Baltimore model and
the cluster disruption parameter derived from our data is in agreement with
the literature.

\subsection{Correlation of Cluster Size with Age and Mass}

The mass of a star cluster is expected to be proportional to the cluster size
since a star cluster would form from a parent molecular cloud once it reaches
a critical density. If we assume a constant density for a cluster, the cluster
size would increase in proportion to $M^{1/3}$. However, it was found by
\citet{lar04} based on the investigation of young star clusters in 18 nearby
spiral galaxies that the correlation between the half light radius $r_{\rm eff}$
and the mass ($M$) of star clusters returned a shallower slope than expected
from a constant-density assumption. For M51 star clusters, \citet{bas05a}
tried to find any correlation between the cluster size and mass but there was
no apparent correlation in their data. More recently, \citet{sch07} also
investigated the size distribution of star clusters using the $HST$ ACS data
and they also reported that they did not find evidence for any direct relation
between mass and radius of the clusters. On the other hand, \citet{lee05}
showed the correlation between the size and mass of star clusters with a
best-fit slope of $0.14 \pm 0.03$, which is in agreement with the result of
\citet{lar04} within $1.5 \sigma$.

\begin{figure}
 \plotone{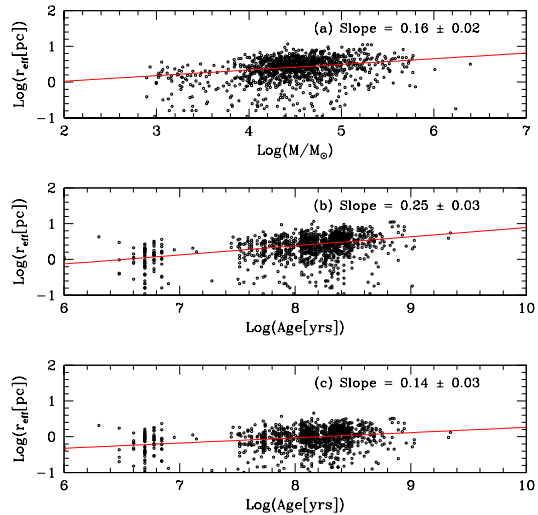}
 \caption{
 (a) Effective radius vs. mass, (b) effective radius vs. age,
 and (c)  effective radius vs. mass corrected for the derived
 size-mass relation of the star clusters in M51 (including both Class 1 and Class 2).
 Solid lines represent the best power law fits (linear fits in the logarithmic scales).
 \label{size}}
\end{figure}

We have investigated the correlation of size with mass of star clusters. The
sizes of star clusters were measured using ISHAPE package \citep{lar99}.
Detailed information on the size measurement and size distribution of M51
star clusters was described in \citet{hwa08}. Figure \ref{size}(a) shows the
distribution of star cluster sizes and masses. The power law relation between
these two parameters is clearly seen in this plot although with large scatters.
We derived the best linear fit from this distribution with a slope of $0.16 \pm
0.02$, which is consistent with the result of \citet{lee05} at the $\sim 1
\sigma$ level. This correlation indicates that the cluster mass distribution in
M51 is shallower than that expected from the constant density assumption.

Figure \ref{size}(b) shows the distribution of star cluster sizes and their ages.
It is seen that there is a trend of cluster size increasing proportional to the
cluster age and this is another result of the `size-of-sample' effect discussed
in Section \ref{disc2}. The best linear fit to the data, shown in a solid line,
gives a slope of $0.25 \pm 0.03$. However, the observed size-age
relationship may simply reflect the fact that the observed mass limit goes up
with age, as shown in Figure \ref{age2mass}. Therefore, in order to check
whether there is an intrinsic trend between star cluster sizes and ages, we
correct the derived star cluster mass-size relation and fit the data again, as
shown in Figure \ref{size}(c). It results in a linear fit with a slope of $0.14
\pm 0.03$, which is steeper than $0.06 \pm 0.02$ by \citet{lee05} and
$0.08 \pm 0.03$ by \citet{sch07}. The relation is shallower than found
before correcting the mass-size relation but it still shows that the size of star
clusters increases with age. This result is in contradiction to that of
\citet{bas05a} in which a slight trend of the decreasing size with increasing
age was shown.

\section{Summary and Conclusion}
\label{sum}

We present the age and mass distribution of about 2,000 star clusters in M51
based on the SED fit of $UBVI$ photometric data. The star cluster age
distribution displays two peaks at $\log t \approx 8.0$ ($100$ Myr) and $\log
t \approx 8.4$ ($250$ Myr). These peaks roughly coincide with the epochs of
dynamical encounters of NGC 5194 and NGC 5195 predicted by a multiple
encounter model \citep{sal00a}. The star cluster formation rate derived from
the age distribution suggests that the cluster formation rate increased
significantly during the period of $100 \sim 250$ Myr ago, when the dynamical
interaction in M51 is expected by the theoretical models
\citep{too72,sal00a}.

The mass function of star clusters can be represented by a single power law,
and the index $\alpha$ appears to change depending on the formation epoch:
$\alpha \approx -2.23 \pm 0.34$ for $t<10$ Myr while $\alpha \approx -1.37
\pm 0.11$ for $t>100$ Myr. Interestingly, the mass function index $\alpha$
changes systematically depending on the formation epochs of star clusters and
the star clusters with $\log t \approx 8.3$ ($200$ Myr) displays the steepest
mass function with $\alpha \approx -1.50$ among star clusters older than
$10$ Myr. This also implies an increased star cluster formation at the
corresponding epoch, possibly incurred by the dynamical interaction of host
galaxies.

The correlation between the maximum mass of star clusters and the
logarithmic age displays a steep slope of about $1.0$ for $\log t<8.2$, while
it is more or less flat for $\log t>8.2$. The age distribution of mass limited
star clusters clearly displays the existence of a plateau over $\log t \approx
7.6$ -- $8.4$ and a bend at $\log t \approx 8.4$. The steep slope of the
maximum cluster mass and the existence of a bend in the mass limited age
distribution indicates that the disruption and evolution of M51 star clusters is
described better with the Utrecht model than with the Baltimore model. The
cluster disruption parameter $\gamma$ derived following the analysis of
\citet{bou03} is $\gamma = 0.53 \pm 0.10$ under the assumption of CIMF
$\alpha = -2.0$, which is in good agreement with the result given by
\citet{lam05}.

We found that the size of star clusters are positively correlated with the mass
of star clusters with a best fit slope $0.14 \pm 0.03$. It shows that the size
of star clusters increases with age even after the correction of the mass-size
relation with a slope $0.16 \pm 0.02$.

\acknowledgements

The authors are grateful to Yanbin Yang for providing the SED fitting program
and acknowledge the support of the BK21 program of the Korean
Government. N.H. was supported in part by Grant-in-Aid for JSPS Fellow No.
20-08325. M.G.L was supported in part by a grant (R01-2007-000-20336-0)
from the Basic Research Program of the Korea Science and Engineering
Foundation. Finally, we thank an anonymous referee for critical and
constructive comments that helped to improve the original manuscript.
\\


\end{document}